\documentclass[a4paper,twocolumn,showpacs5]{revtex4}
\usepackage[english]{babel}
\usepackage{graphicx}
\usepackage{xspace}
\usepackage{mathptmx}

\begin{document}
\title{Epitaxial growth and transport properties of Sr$_2$CrWO$_6$ thin films}

\author{J.~B.~Philipp}\email{Boris.Philipp@wmi.badw.de}

\author{D.~Reisinger}

\author{M.~Schonecke}

\author{M. Opel}

\author{A.~Marx}

\author{A.~Erb}

\author{L.~Alff}

\author{R.~Gross}

\affiliation{Walther-Meissner-Institut, Bayerische Akademie der
Wissenschaften, Walther-Meissner Str.~8, 85748 Garching, Germany}

\date{received, August 21, 2002}


\begin{abstract}
We report on the preparation and characterization of epitaxial thin films
of the double-perovskite Sr$_2$CrWO$_6$ by Pulsed Laser Deposition (PLD).
On substrates with low lattice mismatch like SrTiO$_3$, epitaxial
Sr$_2$CrWO$_6$ films with high crystalline quality can be grown in a
molecular layer-by-layer growth mode. Due to the similar ionic radii of Cr
and W, these elements show no sublattice order. Nevertheless, the measured
Curie temperature is well above 400\,K. Due to the reducing growth
atmosphere required for double perovskites, the SrTiO$_3$ substrate surface
undergoes an insulator-metal transition impeding the separation of thin
film and substrate electric transport properties.
\end{abstract}
\maketitle


Double-perovskites of the form $A_2BB'$O$_6$ are interesting
materials both due to their rich physics and their interesting
properties with respect to applications in spin electronics. They
are half metallic, i.e.~have a high effective spin polarization at
the Fermi level \cite{Kobayashi:98}, and some compounds have a
high Curie temperature $T_C$, e.g. $T_C=635$\,K in Sr$_2$CrReO$_6$
\cite{Kato:02}. Large low field magnetoresistance effects are
found in Sr$_2$FeMoO$_6$ \cite{Kobayashi:98,Garcia:01},
Sr$_2$FeReO$_6$ \cite{Kobayashi:99} and Sr$_2$CrWO$_6$
\cite{Philipp:02}. Thin films of the well studied system
Sr$_2$FeMoO$_6$ have been fabricated by pulsed-laser deposition
(PLD) at relatively high temperatures of about 900$^{\circ}$C
\cite{Manako:99}. However, epitaxial growth was found to be
complicated and difficult to control. There are indications that
high structural quality of films grown on SrTiO$_3$ is associated
with semiconducting behavior \cite{Asano:99,Westerburg:00}. Here,
we report on the epitaxial growth of Sr$_2$CrWO$_6$. Due to the
good lattice match, epitaxial films of this material can be grown
on SrTiO$_3$ substrates in a molecular layer-by-layer growth mode
resulting in high crystalline quality.

\begin{figure}[tb]
\centering{%
\includegraphics [width=0.95\columnwidth,clip=]{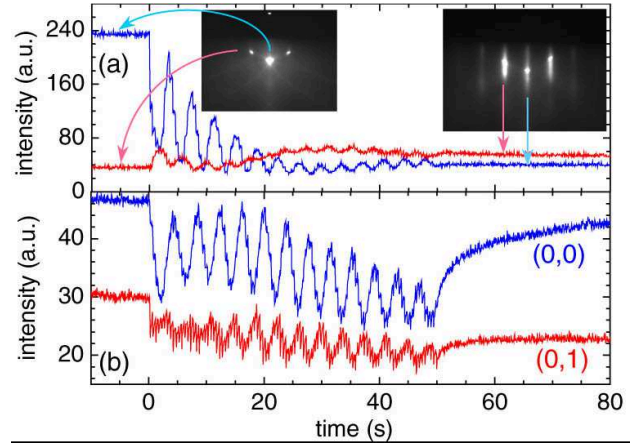}}
 \caption{\small\sf RHEED intensity oscillations observed during
the growth of Sr$_2$CrWO$_6$ on (001) SrTiO$_3$ showing a clear
$\pi$-shift between the (0,0) and (0,1) spot. (a) The first 100 pulses and
(b) the second 100 pulses.  In (b) the intensity of the (0,1) spot is
shifted to lower values for clarity. The inset shows the RHEED patterns
before deposition and after 100 pulses as indicated by the arrows. }
 \label{RHEED}
\end{figure}
\begin{figure}[b]
\centering{%
\includegraphics [width=0.97\columnwidth,clip=]{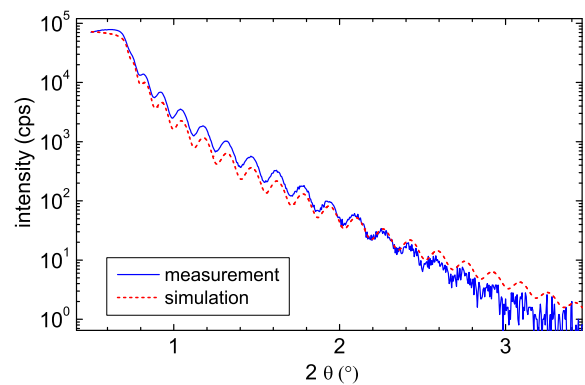}}
 \vspace*{-2mm}\\
 \caption{\small\sf X-ray reflectometry data (solid line) and simulation
data (broken line) \cite{refsim}. }
 \label{Refl}
\end{figure}

Sr$_2$CrWO$_6$ thin films were deposited from stoichiometric targets
\cite{Philipp:02} on atomically  flat, HF etched and annealed (001)
SrTiO$_3$ substrates by PLD using a 248\,nm KrF excimer laser
\cite{Gross:2000a}. Fig.~\ref{RHEED} shows the RHEED (Reflection High
Energy Electron Diffraction) \cite{Klein:99} intensity oscillations of the
$(0,0)$ diffraction spot recorded during the PLD growth of a $c$-axis
oriented Sr$_2$CrWO$_6$ film. The molecular layer-by-layer or
Frank-van~der~Merwe growth mode \cite{Terashima:90} is achieved for a
substrate temperature of 740$^\circ$C, an argon pressure of $2\times
10^{-4}$\,Torr, a laser repetition rate of 2\,Hz, and a laser energy
density on the target of $1.2$\,J/cm$^2$. RHEED was performed with 15\,keV
electrons at an incident angle of about $2^\circ$. To obtain the number of
RHEED oscillations per unit cell, the film thickness has been determined
precisely by X-ray reflectometry (Fig.~\ref{Refl}) and then divided by the
number of observed RHEED oscillations. The derived molecular layer or block
thickness corresponds to half a unit cell ($c/2=4.004\,${\AA}). Note that the
intensity oscillations of the (0,0) and (0,1) spot are out of phase, since
for the (0,1) spot the electrons reflected from different growth planes
interfere constructively (in-Bragg condition). In this case no RHEED
oscillations are expected. The fact that we do observe RHEED oscillations
is caused by multiple and diffuse, incoherent scattering which results in
an increasing (decreasing) intensity with increasing (decreasing) step
density \cite{Korte:97,Reisinger:pre}.

\begin{figure}[tb]
\centering{%
\includegraphics [width=0.95\columnwidth,clip=]{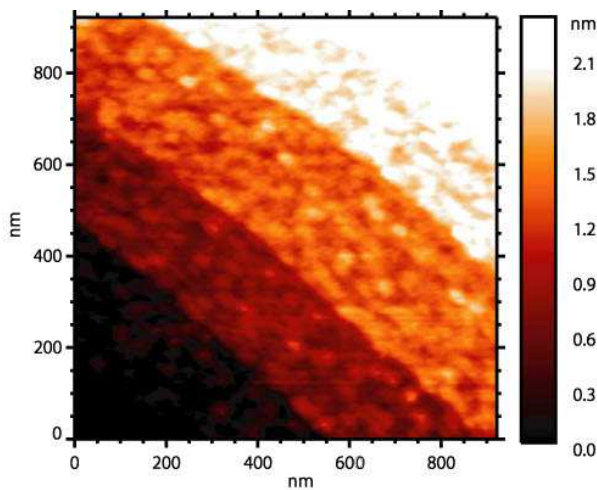}}
 \vspace*{-2mm}\\
 \caption{\small\sf AFM image of a 42\,nm thick epitaxial Sr$_2$CrWO$_6$ film grown on a
HF-etched (001) SrTiO$_3$ substrate}
 \label{AFM}
\end{figure}

Figure \ref{AFM} shows an {\em in situ} AFM picture of a 42\,nm
thick film. Clearly, a terrace structure due to the nonvanishing
substrate miscut with about 4\,{\AA} high steps corresponding to
half a unit cell of Sr$_2$CrWO$_6$ can be seen. The AFM analysis
shows that Sr$_2$CrWO$_6$ grows as an epitaxial film with a
similar crystalline perfection and very smooth surface as the
doped manganites \cite{Gross:2000a,Klein:02}.

\begin{figure}[tb]
\centering{%
\includegraphics [width=0.95\columnwidth,clip=]{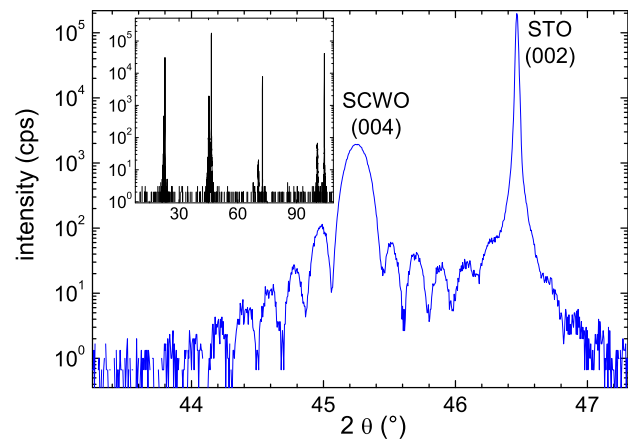}}
 \vspace*{-2mm}\\
 \caption{\small\sf $\theta$-$2\theta$ scan showing the (004) peak
of a Sr$_2$CrWO$_6$ film with the Laue oscillations and the (002) peak of
the SrTiO$_3$ substrate.  The inset shows the $\theta$-$2\theta$ scan from
10$^{\circ}$ to 110$^{\circ}$. }
 \label{XRAY}
\end{figure}

As shown in the inset of Fig.~\ref{XRAY}, within the resolution of our
four circle x-ray diffractometer only (00$\ell$) peaks ($\ell=2,4,6...$)
could be detected. Rocking curves of the (004) peak had a full width at
half maximum (FWHM) of only 0.025$^{\circ}$ which is very close to the
FWHM of the substrate peak. The fact that Laue oscillations were detected
around the film diffraction peaks (Fig.~\ref{XRAY}) indicates that the
strain due to lattice mismatch between film and substrate ($\simeq 0.1$\%
for SrTiO$_3$) is not relaxed, i.e.~the films grow coherently strained. The
unit cell of the high quality strained films has a tetragonal distortion
and is slightly larger than the bulk unit cell ($a_{\rm bulk}=c_{\rm
bulk}=7.813$\,{\AA}, $a_{\rm film}=2\times a_{\rm SrTiO_3}=7.810$\,{\AA} and
$c_{\rm film}=8.005\,${\AA}). Fig.~\ref{Refl} shows X-ray reflectometry data
of a 52\,nm thick Sr$_2$CrWO$_6$ film on a SrTiO$_3$ substrate. Fitting the
data \cite{refsim} gives a film thickness that agrees to that obtained
from counting the RHEED oscillations to within less than $\pm$3\%. Note
that the quality of the fit is extraordinary good.

We have also grown epitaxial Sr$_2$CrWO$_6$ thin films on other substrates
such as NdGaO$_3$, MgAlO$_4$ and LaAlO$_3$. On these substrates a much
rougher surface and a more distorted crystal structure is obtained. This
is caused by the larger lattice mismatch of these substrates to
Sr$_2$CrWO$_6$ as compared to SrTiO$_3$.

An important parameter of the double perovskites is the sublattice
order of the $B$ and $B'$ ions (in our case Cr and W). It is
evident that for perfect order a superstructure is obtained
resulting in a (111) peak in the x-ray diffraction. The intensity
of the (111) peak can therefore be used as a quantitative measure
of the sublattice order of the $B$, $B'$ ions. In our
Sr$_2$CrWO$_6$ thin films {\em no} such peak is found within the
experimental resolution for films grown on (001) or (111)
SrTiO$_3$, (110) NdGaO$_3$, (001) MgAlO$_4$ and (001) LaAlO$_3$.
This gives strong evidence that for our epitaxial Sr$_2$CrWO$_6$
films with high crystalline quality no Cr,W sublattice order is
established. It is likely that this is caused by the very similar
ionic radii of Cr and W resulting in a very small gain of
structural energy by ionic order. We note, however,  that despite
the complete disorder of the $B$ and $B'$ ions a Curie temperature
well above 400\,K is obtained. This raises questions about
theoretical models requiring sublattice order for high $T_C$.  The
saturation magnetization $M_S$ was found to strongly depend on the
growth conditions and the substrate material. A maximum value of
$M_S=1.9\,\mu_B$/f.u at 5\,K was obtained for a 48\,nm thick film
grown on (111) SrTiO$_3$ at $780^\circ$C.

\begin{figure}[tbh]
\centering{%
\includegraphics [width=0.95\columnwidth,clip=]{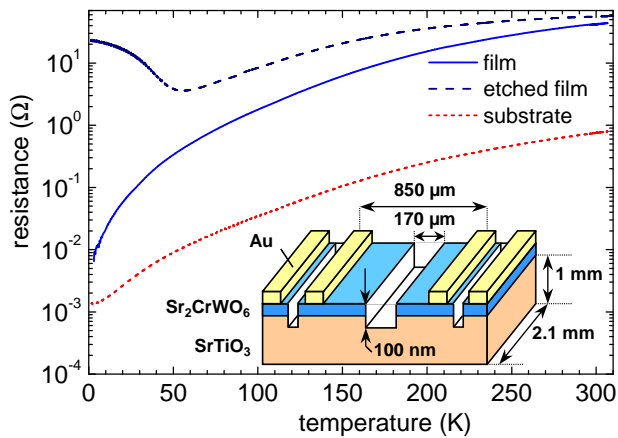}}
 \vspace*{-2mm}\\
 \caption{\small\sf Resistance versus temperature for a 50\,nm thick
Sr$_2$CrWO$_6$ film grown at 740$^{\circ}$C (solid line). The
inset shows the sample geometry.  The dashed line represents the
$R(T)$ curve of the same film after etching a 100\,nm deep and
170\,$\mu$m wide trench to interrupt the Sr$_2$CrWO$_6$ film. As a
reference, the dotted curve shows the $R(T)$ curve of a SrTiO$_3$
substrate annealed at 820$^{\circ}$C in Ar.}
 \label{RT}
\end{figure}

Resistance vs.~temperature was measured with a standard four probe
configuration. All films not grown on SrTiO$_3$ substrates show
semiconducting behavior what might be attributed to the large lattice
mismatch. We note however, the double perovskite films like Sr$_2$CrWO$_6$
or Sr$_2$FeMo$_6$ are grown under very low oxygen partial pressure
\cite{Manako:99,Westerburg:00}, vacuum \cite{Asano:99} or Argon atmosphere
\cite{Yin:99,Westerburg:00,Philipp:02} in a temperature range between 700
and 900$^{\circ}$C. Under these conditions (e.g.~$T=800^{\circ}$C, $p_{\rm
O_2}=1 \times 10^{-8}$\,Torr) the surface of the SrTiO$_3$ substrates
becomes oxygen deficient within a few minutes. That is, a few $\mu$m
thick, oxygen deficient surface layer (SrTiO$_{3-\delta}$) is obtained
resulting in a conducting substrate surface \cite{Szot:02}. This process
is further supported by the reducing plasma during the PLD process. We
found that the resistance of the SrTiO$_3$ substrate depends strongly on
temperature, reducing atmosphere and time. In Fig.~\ref{RT}, the $R(T)$
curve of a SrTiO$_3$ substrate annealed at 820$^{\circ}$ in Ar is shown as a
reference. After the treatment the substrate turns black and shows
metallic behavior.

Fig.~\ref{RT} also shows the $R(T)$ curve of a Sr$_2$CrWO$_6$ film
grown at a low substrate temperature of only 740$^{\circ}$C.
Evidently, whether or not the metallic behavior results from the
film or the substrate, we have etched trenches into the film (see
inset in Fig.~\ref{RT}) interrupting the film and forcing the
measurement current to flow in parts through the SrTiO$_3$
substrate. As can be seen in Fig.~\ref{RT} the resistance is still
metallic above 50\,K but has been increased significantly by
roughly one order of magnitude. However, this difference cannot be
unambiguously attributed to the interruption of the Sr$_2$CrWO$_6$
film alone because the trenches are deeper than the film
thickness, i.e.~the outermost layers of the SrTiO$_3$ which may be
well conducting are also cut from the current path. Hence, for the
Sr$_2$CrWO$_6$ films on SrTiO$_3$ we only have some indication
that these films are metallic, however, we cannot definitely prove
it at present. Due to the problems of conducting substrate
surfaces, in Fig.~\ref{RT} we plot the resistance of the sample
configuration given in the inset and do not plot the resistivity.

In summary, we have grown epitaxial Sr$_2$CrWO$_6$ thin films with very
high crystalline quality on SrTiO$_3$ by PLD. Since the ionic radii of Cr
and W are almost identical, no sublattice order of these elements is
achieved. Nevertheless, we observe Curie temperatures well above 400\,K
for these disordered films. The transport properties of the Sr$_2$CrWO$_6$
films on SrTiO$_3$ could not unambiguously be separated from those of the
substrate, since the SrTiO$_3$ substrate is turned metallic under the
reducing deposition conditions.

This work was supported by the Deutsche Forschungsgemeinschaft
(project Al/560) and the BMBF (project 13N8279).


\end{document}